\documentclass{iopconfser}

\usepackage[utf8]{inputenc}
\usepackage{braket,mathrsfs,mathtools,slashed}
\let\originalleft\left
\let\originalright\right
\renewcommand{\left}{\mathopen{}\mathclose\bgroup\originalleft}
\renewcommand{\right}{\aftergroup\egroup\originalright}

\usepackage[normalem]{ulem}
\usepackage[locale = US]{siunitx} 
\DeclareSIUnit\annum{a}
\usepackage{dsfont}
\usepackage{textgreek}
\usepackage{placeins,graphicx,graphics,multirow,tabularx} 
\usepackage{subcaption}
\usepackage{nccmath} 
\usepackage{footmisc} 
\usepackage{hyperref}  
\usepackage{enumitem} 
\usepackage{tikz} 
\usepackage{tikzsymbols}
\usepackage{appendix}
\usepackage[newzealand]{babel}
\usepackage{microtype}
\usepackage[T1]{fontenc} 
\usepackage{csquotes} 

\definecolor{lime}{HTML}{A6CE39} 
\definecolor{darkgreen}{rgb}{.125,.5,.25}
\newcommand{\orcidicon}{%
	\begin{tikzpicture}
		\draw[lime, fill=lime] (0,0) 
		circle [radius=0.16] 
		node[white] {{\fontfamily{qag}\selectfont \tiny ID}};
		\draw[white, fill=white] (-0.0625,0.095) 
		circle [radius=0.007];
	\end{tikzpicture}
	\hspace{-3mm}
}
\newcommand\orcidMatt{{\href{https://orcid.org/0000-0003-1088-6485}{\orcidicon}}}
\newcommand\orcidSeSc{{\href{https://orcid.org/0000-0003-1997-0026}{\orcidicon}}}
\newcommand\orcidJess{{\href{https://orcid.org/0000-0002-2669-2899}{\orcidicon}}}
\newcommand\orcidJCFe{{\href{https://orcid.org/0000-0003-2441-5801}{\orcidicon}}}

\newcommand{\defi}{\mathrel{\mathop:}=}

\newcommand{\dif}{\ensuremath{\operatorname{d}}\!}

\newcommand{\LambdaCDM}{{\textLambda\kern-0.06667em CDM}}

\begin{document}

\title{Primordial Black Holes, Charge, and Dark Matter: Rethinking Evaporation Limits}

\author{Sebastian Schuster$^{1}$ \orcidSeSc, Jessica Santiago$^{2}$ \orcidJess, Justin Feng$^{3}$ \orcidJCFe and Matt Visser$^{4}$ \orcidMatt}

\affil{$^1$Fysikum, Stockholm universitet, Stockholm, Sweden}
\affil{$^2$Leung Center for Cosmology \& Particle Astrophysics,	National Taiwan University, Taipei, Taiwan}
\affil{$^3$Central European Institute for Cosmology and Fundamental Physics, Czech Academy of Sciences, Prague, Czech~Republic}
\affil{$^4$School of Mathematics and Statistics, Victoria University of Wellington, New~Zealand}

\email{\textsuperscript{1}\url{sebastian.schuster@fysik.su.se}, \textsuperscript{2}\url{jessica.santiago@cpt.univ-mrs.fr}, \textsuperscript{3}\url{feng@fzu.cz}, \textsuperscript{4}\url{matt.visser@sms.vuw.ac.nz}}

\begin{abstract}
	Limits on the dark matter fraction of small mass primordial black holes from Hawking radiation are predominantly derived from the assumption of a Schwarzschild black hole evaporating. However, astrophysical black holes are usually much more realistically modelled by the rotating Kerr black hole solution. Meanwhile, electromagnetically charged black holes are astrophysically of little importance due to their fast neutralisation in the present universe. Dark matter is not just a possible solution to issues of astrophysics and cosmology, but also to issues of the standard model of particle physics. Extensions of this model thus can lead to charges present in the early universe which remain preserved in the charge of primordial black holes --- even when the corresponding particles have disappeared from the particle content of the present epoch of the universe. Here, we report on a thorough proof-of-concept that such charges can greatly change evaporation limits for primordial black hole dark matter. Special emphasis is placed on (near-)extremal black holes, for which this effect is especially pronounced.
\end{abstract}

The cosmological standard model, \LambdaCDM, contains with the dark sector of dark energy and dark matter two of the biggest riddles to be solved by current observational and experimental efforts spanning astronomy and particle physics. Broadly speaking, to reconcile our understanding of general relativistic cosmology with the observed matter distribution and behaviour, some form of dark energy and dark matter is needed. In this article, we will concern ourselves with the latter. Dark matter can, again broadly speaking, be lumped into either weakly interacting massive particles (\enquote{WIMPs}) on the one hand, and massive compact halo objects (\enquote{MACHOs}) on the other. The former are modifications of the standard model of particle physics, the latter unseen objects large and massive enough to be governed by gravitational physics first and foremost. These are not the only options: at the very least, one could also opt to change the gravitational laws or cosmological models---we will not do so here. Usually, phenomenological studies have to focus on one or the other models of dark matter. Beyond cosmology itself, however, particle physics itself has puzzles that additional particle content could solve, and this particle content in turn would appear as dark matter in the astrophysical and cosmological context.

In a recent article~\cite{Santiago:2025rzb}, the present authors studied in a simple, proof-of-concept model how such a extension of the standard model of particle physics can modify existing limits on the fraction $(f_{\textrm{PBH}})$ of dark matter made up of low-mass, primordial black holes, see also \cite{Baker:2025zxm}. Crucially, most of the low-mass bounds for such primordial black holes as dark matter are evaporation bounds: Too light black holes would have evaporated by now through the Hawking process, or at the very least be visible as hot sources of radiation. Concretely, the lack of direct or indirect observation of evaporating black holes effectively rules out primordial black holes as dark matter if their mass $M_{\textrm{PBH}}$ is below $\sim 10^{-15}M_{\odot}$ ~\cite{Carr:2020gox,Luo:2020dlg, Keith:2020jww, Korwar:2023kpy, Boudaud:2018hqb, Auffinger:2022khh, HESS:2023zzd}.

However, the fine-print in these bounds is important. Most crucially, the vast majority of extant studies is built on the evaporation of Schwarzschild black holes, \emph{i.e.}, non-rotating, uncharged, spherically symmetric black holes. Astrophysically, it is to be expected that primordial black holes would at the very least also have angular momentum, giving the only axisymmetric Kerr black hole as a more accurate model. This already lowers their temperature, but also complicates the analysis as greybody factors and superradiance play an increasing role~\cite{Page:1976ki,Brito:2015oca}. We will keep the simplicity of spherical symmetry, but focus on a different type of complication: 
Changes to the bounds on $(f_{\textrm{PBH}})$ for low masses, if charge is allowed as additional parameter.

Concretely, this allows for near-extremal black holes with a near-zero Hawking temperature. This greatly increases the black hole's lifetime. However, this only works if the black hole's charge $Q$ is not due to standard electromagnetism. If the charge was electromagnetic in nature, the black hole would neutralize very quickly through accretion~\cite{Eardley:1975kp,Gong:2019aqa}, not to mention very observable effects even for small residual charges far from extremality~\cite{Zajacek:2019kla}. Instead, we will consider a very simple extension of the standard model of particle physics: We add a dark $U(1)$, for all intents and purposes described by (quantum) electrodynamics. The key features to note are: (1) Two new free parameters, the charge $e_\chi$ and mass $m_\chi$ of the corresponding \enquote{dark} electron, (2) this matter content does not couple to (currently present) standard model particle physics, and (3) while present in the early universe when the primordial black holes were formed (and thus could acquire dark charge $Q$), dark (anti-)electrons are not present in the present epoch of the universe anymore and cannot neutralize the black hole through accretion. While this is primarily meant as a proof-of-concept, such an additional $U(1)$ is not fully ruled out itself~\cite{Ackerman:2008kmp, Feng:2009mn, Agrawal:2016quu}. In our analysis~\cite{Santiago:2025rzb}, we used the model of Hiscock and Weems~\cite{Hiscock:1990ex}, in the following referred to as \enquote{HW}.

\begin{figure}[ht]
	\begin{subfigure}{.33\textwidth}
		\includegraphics[width=\textwidth]{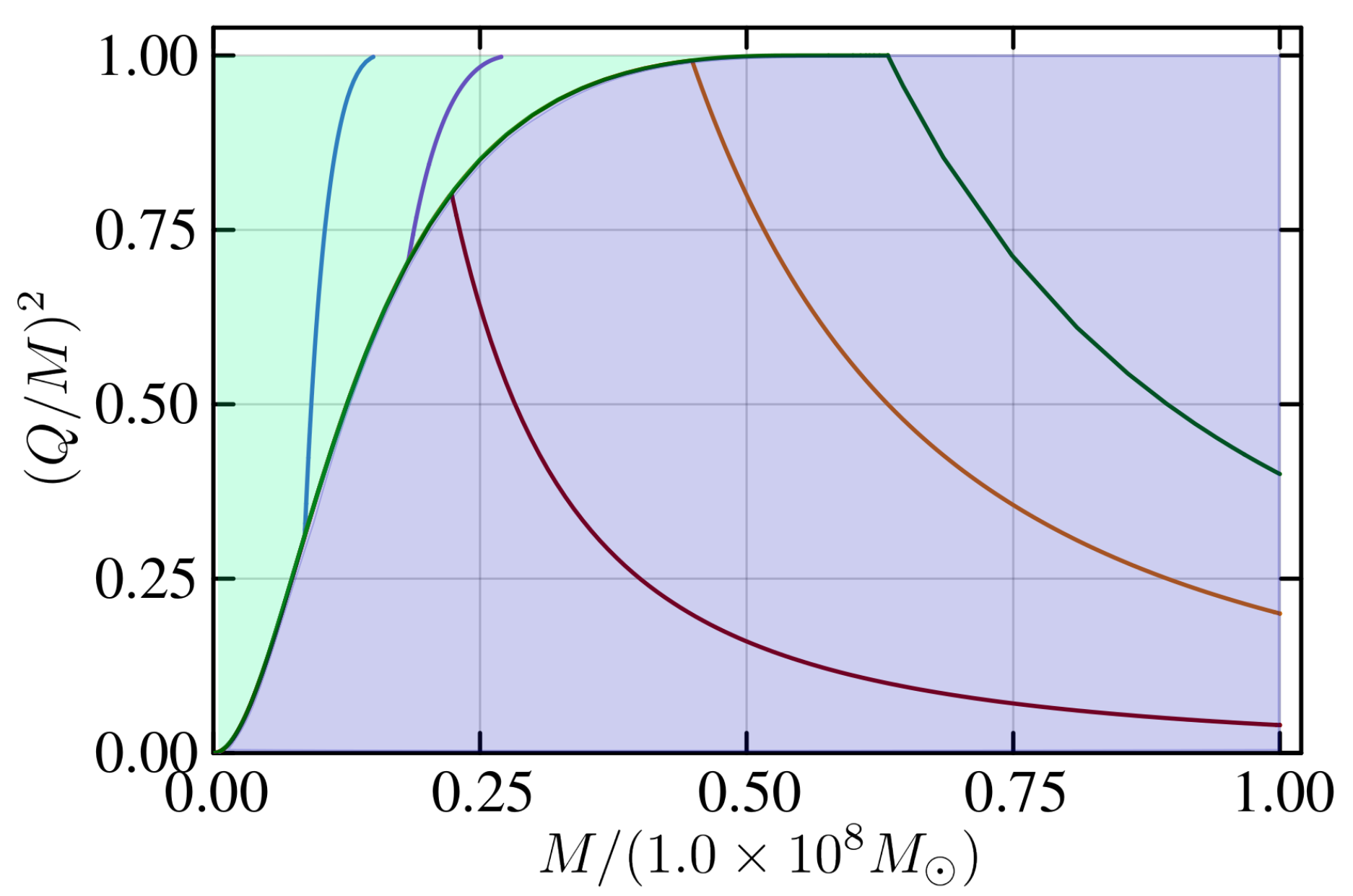}
		\caption{}
		\label{fig:HW-generic}
	\end{subfigure}
	\begin{subfigure}{.33\textwidth}
		\includegraphics[width=\textwidth]{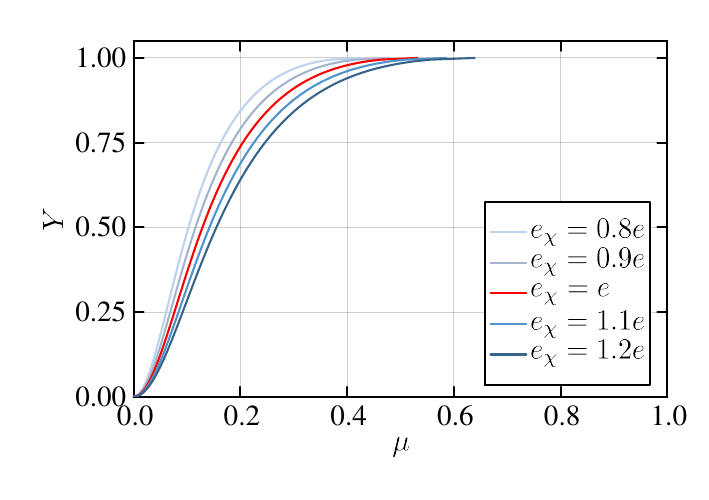}
		\caption{}
		\label{fig:HW-vary-echi}
	\end{subfigure}
	\begin{subfigure}{.33\textwidth}
		\includegraphics[width=\textwidth]{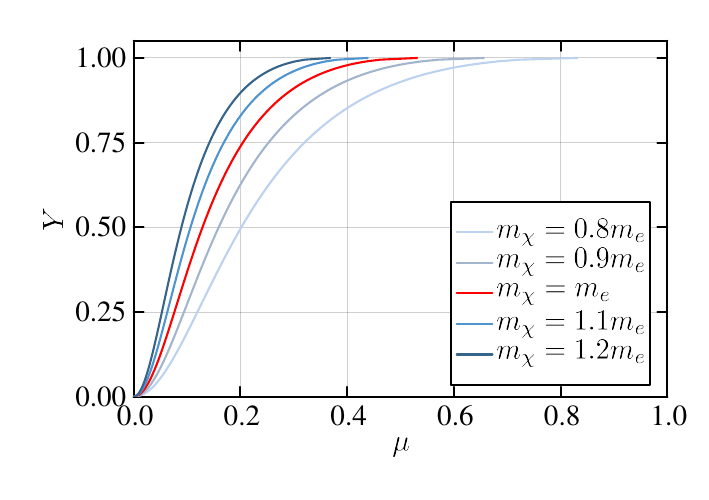}
		\caption{}
		\label{fig:HW-vary-mchi}
	\end{subfigure}
	\caption{The general features of the HW evaporation model, and its dependence on the dark QED parameters $e_\chi$ and $m_\chi$. Figure (a) summarizes the three core features of a black hole's trajectory in the parameter space $(Y=Q^2/M^2, M = \mu M_\text{s})$, where $M_\text{s}$ is chosen such that $\mu\in[0,1]$ for the chosen parameter range: (1) Starting parameters in the light green region are dominated by charge loss and the Schwinger effect. (2) Starting parameters in the light blue region are dominated by mass loss and the Hawking effect. (3) Trajectories approach an attractor curve separating the two regions. Figures (b) and (c) show how a semi-analytic approximation of this attractor curve (derived in~\cite{Santiago:2025rzb}) depends on the dark electron's charge $e_\chi$ and mass $m_\chi$, respectively.}
	\label{fig:HW-summary}
\end{figure}
The HW model adapted to dark $U(1)$ physics are two non-dimensionalized, coupled ordinary differential equations in terms of $Y=Q^2/M^2$ and $M = \mu M_\text{s}$, with $M_\text{s}$ chosen such that $\mu\in[0,1]$. Concretely,
\begin{equation}
	\frac{\dif \mu}{\dif \tau} 	= -\frac{ \left(H(\mu,Y)+S(\mu,Y) Y^2\right)}{ (\sqrt{1-Y}+1)^4},\quad
	\frac{\dif Y}{\dif \tau} 	= \frac{2 \left( H(\mu,Y) -S(\mu,Y) \left(1-Y+\sqrt{1-Y}\right)Y\right)Y}{\mu (\sqrt{1-Y}+1)^4},\label{eq:HWnew}
\end{equation}
where
\begin{equation}
	H(\mu,Y) \defi\frac{\left(\sqrt{9-8 Y}+3\right)^4 (1-Y)^2}{\mu ^2 (\sqrt{1-Y}+1)^4 (3-2 Y+\sqrt{9-8 Y})}, \quad
	S(\mu,Y) \defi
	\exp\left\{b_0\left[z_0-{\mu \left(\sqrt{1-Y}+1\right)^2}/{\sqrt{Y}}\right]\right\}.
	\label{eq:H-and-S-eq}
\end{equation}
Here, the dimensionless constants $s_0$, $z_0$, and $b_0$ are given by:
	\begin{equation}
		s_0 =\frac{\alpha  \hbar }{1920 \pi  M_\textrm{s}^2}
		, \quad
		z_0 =\frac{e_\chi \hbar}{\pi m_\chi^2 M_\textrm{s}}\ln\left(\frac{960 e_\chi^4 M_\textrm{s}^2}{\pi ^2 \alpha m_\chi^2 \hbar ^2}\right)
		,\quad
		b_0 =\frac{m_\chi^2 M_\textrm{s} \pi}{e_\chi \hbar}.
	\end{equation}
Finally, $\alpha\in \{\num{0.2679}, \num{2.0228}\}$ is a numerical parameter to correctly account for the scattering cross-section of all massless particles considered in the model, see \cite{Santiago:2025rzb,Hiscock:1990ex} for details. $M_\textrm{s}$ is a scaling mass used to confine the considered, non-dimensionalized mass range to the interval $[0,1]$. The evolution is assuming the following three conditions, involving the characteristic charge scale for the Schwinger effect, $Q_0=\hbar e_\chi/(\pi m_\chi^2)$: 
\begin{equation}
	\textrm{(i)}\;\; M \gg \frac{\hbar}{m_\chi}, \quad 
	\textrm{(ii)}\;\; \frac{e_\chi^3 Q}{m_\chi^2 r^2} \ll 1, \textrm{~~and~~} \textrm{(iii)}\;\;r_+^2 \gg QQ_0.
	\label{eq:summ_conditions}
\end{equation}
For more detailed discussion of these conditions for dark electromagnetism and scaling relations useful for their analysis, we refer the reader again to \cite{Santiago:2025rzb}, section IV.C.

The general behaviour of the evolution equations~\eqref{eq:HWnew} is depicted in figures~\ref{fig:HW-summary}. Concretely, one has a charge dissipation zone for low black hole mass $M$ and high black hole charge $Q$ divided from a mass dissipation zone by an attractor curve, see figure~\ref{fig:HW-generic}. In reference~\cite{Santiago:2025rzb}, we derived a semi-analytic expression of high accuracy for this attractor curve, beginning at a non-dimensional mass scale $z_0$. The dependence of this approximated attractor curve on $e_\chi$ and $m_\chi$ is shown in figures~\ref{fig:HW-vary-echi} and~\ref{fig:HW-vary-mchi}.
\begin{figure}[ht]
	\begin{subfigure}{.5\textwidth}
		\includegraphics[width=\textwidth]{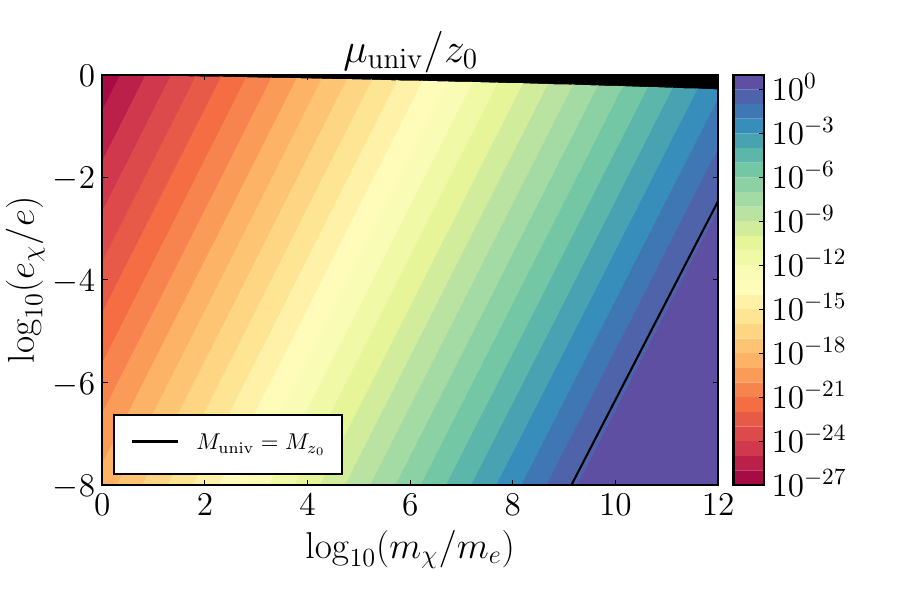}
		\caption{}
		\label{fig:muz-contour}
	\end{subfigure}
	\begin{subfigure}{.5\textwidth}
		\includegraphics[width=\textwidth]{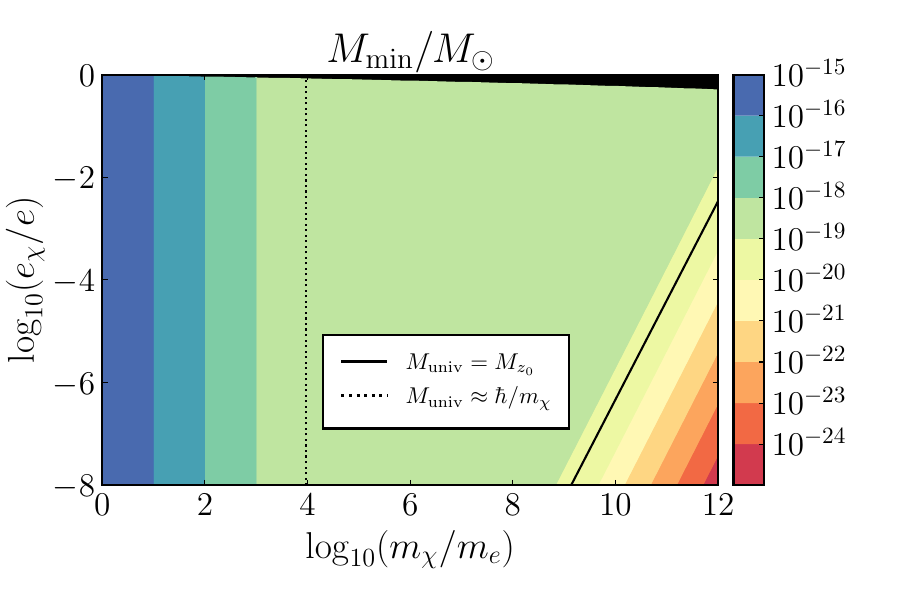}
		\caption{}
		\label{fig:Muniv-contour}
	\end{subfigure}
	\caption{The minimal mass for a charged black hole to have a lifetime of at least the age of the universe and its dependence on $e_\chi$ and $m_\chi$. The black region at the top is ruled out by condition {\eqref{eq:summ_conditions}\babelhyphen{nobreak}(ii)}. $M_{z_0}$ is the mass scale corresponding to $\mu=z_0$, where the approximated attractor curve reaches $Y=1$. On the left, this is shown in non-dimensionalized mass units $\mu_\textrm{univ}$, on the right, in solar masses $M_\odot$. The dotted line in figure~\ref{fig:Muniv-contour} corresponds to the limit of conditon~\eqref{eq:summ_conditions}\babelhyphen{nobreak}(i).}
	\label{fig:summary}
\end{figure}

The trajectory of a (sufficiently) charged black hole through parameter space takes significantly more time than that of a similarly massive Schwarzschild black hole. This can be seen in figures~\ref{fig:summary}, with the upshot being that charged black holes easily avoid evaporation bounds for uncharged black holes.

As the aim is partly a proof of concept, we did not study formation of such black holes, beyond the assumptions for how they could have survived until now mentioned above. In \cite{Bai:2019zcd}, a brief discussion for such formation can be found. The large, extant literature on formation of \emph{uncharged} black holes should be sufficient evidence that charged black hole formation needs more study, too. Likewise, this discussion does not stop at Reissner--Nordström black holes: No-hair theorems do not apply to non-vacuum situations \cite{Visser:1992qh} (as the early universe is wont to be), non-abelian gauge theories, or modifications of gravity \cite{Alonso-Monsalve:2023brx}. Naturally, for different gauge fields, the Schwinger effect itself would have to be adapted. Also, in even more extreme black holes than considered in our work, phase space considerations further suppress evaporation, and thus extend the black holes' lifetime even further \cite{Page:2000dk,Brown:2024ajk}, not to mention more radical proposals \cite{Montefalcone:2025akm}.

Overall, all that remains is to reiterate that accurate evaporation limits for PBH mass ranges need to account for different particle physics effects and black hole metrics than just the Hawking evaporation of a Schwarzschild black hole. Other radiation channels and black hole parameters can together significantly open up low mass windows, and thus greatly change the PBH dark matter fraction.

\section*{Acknowledgements}
JS and JF acknowledge support from the Taiwan National Science and Technology Council, grant No. 112-2811-M-002-132. JF also acknowledges support from the European Union and Czech Ministry of Education, Youth and Sports through the FORTE project No. CZ.02.01.01/00/22\_008/0004632. SS is supported by a stipend for foreign postdoctoral researchers in Sweden of the Wenner Gren Foundations. 
MV was supported, during early phases of this project, by the Marsden Fund, via a grant administered by the Royal Society of New Zealand.

The authors wish to thank Lucas Cornetta and Anne Reinarz for their help with the numerics during the early stages of the project. 
The authors also wish to thank Will Barker, Harry Goodhew, Emma Greenbank, Joaquim Iguaz Juan, David Kaiser, Gordon Lee, Phillip Levin, Ilia Musco, Don Page, Martin Spinrath, and Jan Tristram Acuña for helpful discussions and suggesting references.

\providecommand{\newblock}{}

\end{document}